\documentclass[%
superscriptaddress,
preprint,
 amsmath,amssymb,
 showkeys,
]{revtex4-2}

\usepackage{graphicx}% Include figure files
\usepackage{dcolumn}% Align table columns on decimal point
\usepackage{bm}% bold math
\begin{document}

%\preprint{APS/123-QED}

\title{Laser pulse-droplet interaction enables the deformation and fragmentation of droplet array}% Force line breaks with \\

\author{D. Chaitanya Kumar Rao }
\thanks{These two authors contributed equally}
 \affiliation { Department of Mechanical Engineering, Indian Institute of Science, Bangalore 560012, India}%
\affiliation { Department of Physics, University of Gothenburg, SE-412 96 Gothenburg, Sweden}
\author{Awanish Pratap Singh}
\thanks{These two authors contributed equally}

\author{Saptarshi Basu}
 \email{sbasu@iisc.ac.in}
 \affiliation { Department of Mechanical Engineering, Indian Institute of Science, Bangalore 560012, India}%


\begin{abstract}
 Droplet-droplet interactions is ubiquitous in various applications ranging from medical diagnostics to enhancing and optimizing liquid jet propulsion. We employ an experimental technique where the laser pulse interacts with a micron-sized droplet and causes optical breakdown. The synergy of a nanosecond laser pulse and an isolated spherical droplet is accurately controlled and manipulated to influence the deformation and fragmentation of an array of droplets. We elucidate how the fluid dynamic response (such as drop-drop and shock-drop interactions) of an arrangement of droplets can be regulated and optimally shaped by laser pulse energy and its interplay with the optical density of liquid target. A new butterfly type breakup is revealed, which is found to result in controlled and efficient fragmentation of the outer droplets in an array. The spatio-temporal characteristics of a laser-induced breakdown dictate how shock wave and central droplet fragments can influence outer droplets. The incident laser energy and pulse width employed in this work are representative of diverse industrial applications such as surface cleaning, nano-lithography, microelectronics, and medical procedures such as intraocular microsurgery.
\end{abstract}

%Use showkeys class option if keyword
\maketitle

%\tableofcontents
\section{\label{sec:Intro}Introduction}
Effective liquid atomization is crucial in a broad spectrum of industrial applications such as combustion engines, spray drying, and medical nebulizers \cite{Villermaux_2007, Yarin_2006,Villermaux_2020}. For instance, the fuel atomization characteristics (such as drop-drop interactions) in automotive and gas turbine engines are essential in ascertaining combustion stability, efficiency, and pollutant emissions. To evaluate these vital parameters, understanding the spatio-temporal distribution of droplets that originate from the breakup process is essential \cite{Lefebvre_1988,Lefebvre_2010,Villermaux_2007}. Considering that droplets constitute the sub-grid level of a spray, it is important to scrutinize the deformation mechanisms and atomization pertaining to a single isolated droplet \cite{Lyu_2021,Rao_2017,Rao_2018}.

Fragmentation of liquid can be achieved by a focused laser pulse. The focusing of a high-intensity laser pulse within a fluid beyond a specific threshold ($\sim$ GW/cm$^{2}$) leads to optical breakdown \cite{PhysRevLett.11.401,RAMSDEN_1964,Morgan_1975}. The breakdown is observed as a luminous spark with intense sound, accompanied by the creation of charged particles while the absorption of laser pulse by the liquid increases. The breakdown leads to the formation of high-temperature plasma of the order of 10$^{6}\:^{o}$K \cite{Morgan_1975}. The rapid expansion of plasma is followed by the generation of a spherical shock wave that grows with time. 

Laser-induced breakdown (LIB) inside medium such as liquid droplets provide a prospect of spatially isolated energy deposition that cannot be accomplished with any other optical method. The development of well-established ultra-short pulse lasers has made it feasible to achieve spatial and temporal accuracy in wide ranging applications such as nano-lithography \cite{Torretti_2020,Klein_2020}, surface cleaning \cite{Ohl_2006,Jang_2009,Jang_2011,Lee_2001}, combustion \cite{Singh_2019,Bradley_2004}, and microsurgery \cite{Vogel_2003,Vogel_2005}. 

Laser shock cleaning process based on LIB has also demonstrated the ability to remove nanoscale particles ($\sim$ 10 nm) from silicon wafers, and it has applications in semiconductor industry \cite{Lee_2001}. These micro/nano-particle contaminations alter precision manufacturing quality in the semiconductor industry and high-threshold optical components. However, the conventional cleaning techniques do not function efficiently at micro-and nano-length scales \cite{Lee_2001}. Laser-induced breakup of a micron-sized droplet can produce a high-speed jet with velocities of the order of 1000 m/s. It was demonstrated that the impingement of the atomized droplets on the contaminated substrates could efficiently remove the nanoparticles under hydrodynamic drag forces \cite{Ahn_2013}. The distinctive attribute of LIB allows its applicability in biological and biomedical applications \cite{Vogel_1997,Palczewska_2020,Vogel_2003,Vogel_2005}. The laser-induced optical breakdown has been employed in sophisticated non-intrusive intraocular microsurgery with nanosecond to femtosecond laser pulse \cite{Vogel_1995,Vogel_2003, Vogel_2005}. 

The focusing of a laser pulse within a droplet can result in cavitation, deformation, or fragmentation of a freely falling \cite{Gelderblom_2016,Klein_2020} or acoustically levitated \cite{Gonzalez_Avila_2016,Zeng_2018} droplets. 
Gelderblom et al. \cite{Gelderblom_2016} investigated the deformation while Klien et al. \cite{Klein_2020} studied the fragmentation of freely falling droplets with a nanosecond pulse laser. The significance of shape of laser pulse and droplet properties on the droplet deformation and breakup was examined. %It was found that a tightly focused laser pulse leads to the development of a curved sheet with maximum lateral expansion. In contrast, a less focused pulse culminates in a comparatively smaller expansion with a flat-symmetrical sheet \cite{Gelderblom_2016}. Furthermore, it has been reported that the breakup of the sheet is dictated by the Weber number (based on the propulsion speed of the droplet), kinetic energy partition between propulsion and sheet expansion, and the amplitude of the corrugations present during the initial droplet acceleration \cite{Klein_2020}. 
Avila and Ohl \cite{Gonzalez_Avila_2016} investigated the deformation, breakup, and shock dynamics corresponding to acoustically levitated droplets of different sizes. Based on the droplet size and laser energy, distinct breakup modes (from course fragmentation to rapid atomization) have been reported.
Although a number of studies have been conducted on the interaction of a laser-pulse with a single droplet, the literature lacks any study on laser-pulse interaction with an array leading to drop-drop or shock-drop interactions. Moreover, the direct interaction of the laser pulse with a droplet is not always beneficial, especially when the target surfaces are photosensitive (e.g. eye).

In this paper, we analyze the breakup of an array of acoustically levitated droplets with a nanosecond pulse laser. We report the breakup dynamics of water and Diesel droplets upon their interaction with a laser pulse and their subsequent collision (droplet fragments and shock wave) with the neighboring droplets. Furthermore, we have attempted to elucidate questions such as how the fragmentation modes (based on laser energy and droplet properties) of central droplets affect the degree of deformation and breakup of neighboring droplets? Or what role mechanisms such as ligament-mediated breakup play in the narrow or broader size distribution of the secondary droplets?

\section{Materials and Methods}
The droplet arrays are vertically arranged with three different droplet sizes ranging from 200 $\mu$m to 600 $\mu$m. The mono-disperse droplets with distinct physico-chemical properties (water and Diesel) are levitated on the pressure nodes of a single-axis levitator (tec5) with 100 kHz frequency. Water is used in the present work since it has been established that the optical breakdown in distilled water is comparable to the breakdown thresholds in ocular medium \cite{docchio1986experimental,Vogel_2008}. Furthermore, the mechanical properties of water are similar to those of an aqueous tumor and the vitreous substance \cite{Vogel_1999}. Similarly, Diesel is used as a representative practical liquid that is widely used not only in internal combustion and gas turbine engines but also in the manufacturing, construction, and oil and gas industries \cite{Lefebvre_2010, Lefebvre_1988}.
\begin{figure}%[tbhp]
\centering
\includegraphics[width=0.85\linewidth]{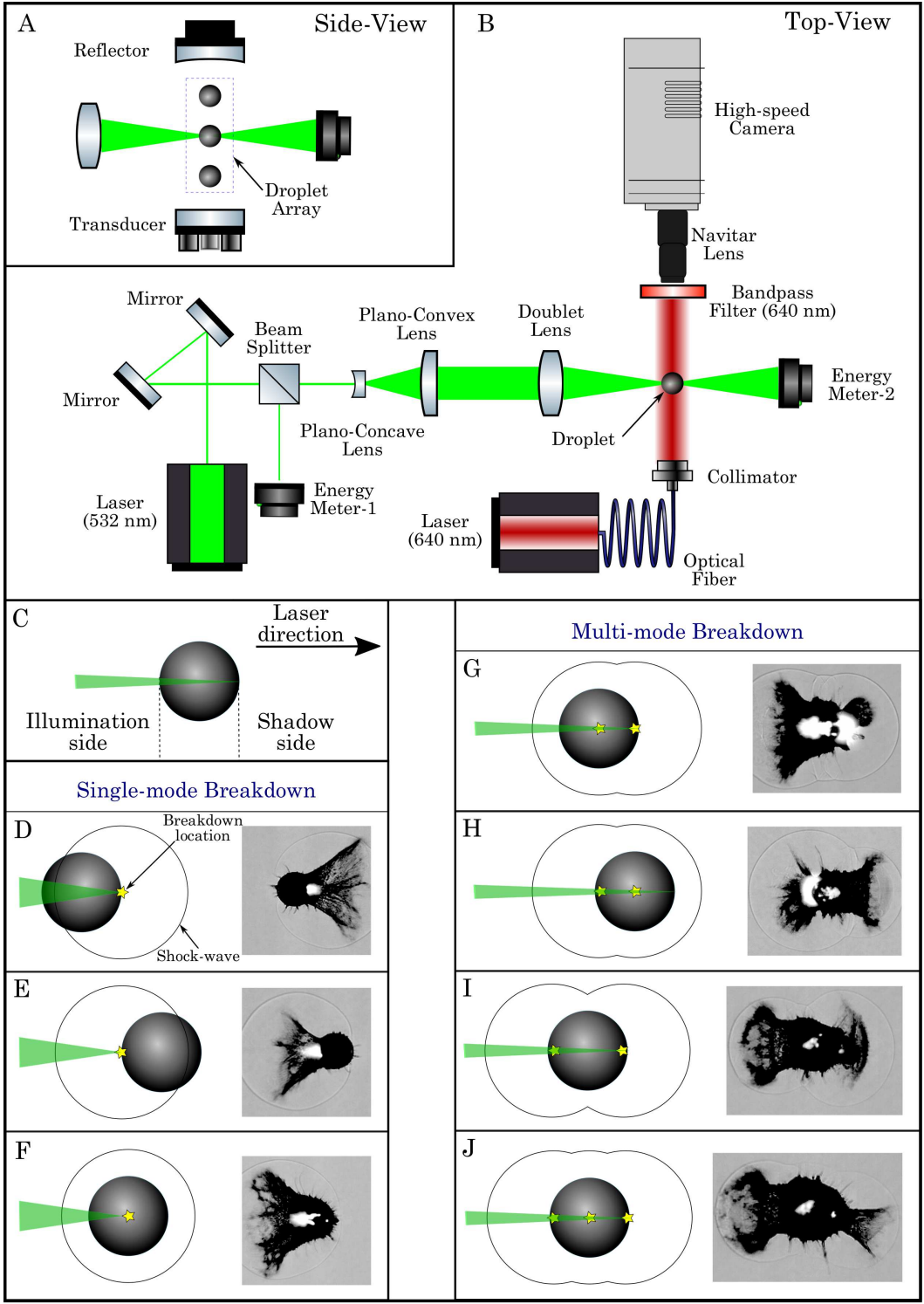}
\caption{Schematic representing (A) Side view (displaying acoustically levitated droplets) and (B) top view (displaying optical arrangement and visualization method) of the experimental setup. (C) Representation of shadow side and illumination side with respect to laser direction and droplet location. (D-J) Breakdown zones and corresponding breakup modes with respect to laser direction and droplet location.}
\label{fig:fig4}
\end{figure}

A Q-switched Nd: YAG laser with a wavelength of 532 nm (InnoLas Spitlight 400) was used to create the laser-induced breakdown to fragment the central droplet. The laser beam with a Gaussian spatial profile was expanded using a Galilean beam expander (approximately 25.5 mm) and focused via a laser-grade achromatic doublet lens ($f =$ 150 mm). The laser delivered a pulse duration of 6 ns with incident energies ranging from 5 mJ to 80 mJ and was used in a single pulse mode. The beam waist was calculated to be $\sim$ 3 $\mu$m following Singh et al. \cite{Singh_2019}. Laser pulses were aimed precisely at the central droplet that was placed near the focal point ( $f \pm $ 1600 $\mu$m). The droplet location was chosen such that different modes of breakdown are observed. The laser pulse energies in front of and behind the breakdown location were measured using a pyroelectric sensor.

Figure \ref{fig:fig4}A \& \ref{fig:fig4}B display the schematic diagram of the experimental apparatus. A high-speed camera (Photron SA5) and a high-speed laser for illumination (CAVILUX® Smart UHS, 640 nm) were utilized to investigate the droplet breakup and the expansion of shock waves generated by the breakdown process. The high-speed images were recorded at 40 kHz and the single shot images (with laser illumination) were acquired at 50 ns interval to capture the shock waves and the subsequent evolution of the droplet fragments. The spatial resolution of recorded images was 9.98 $\mu$m/pixel. A digital delay generator (BNC 745T) was used to synchronize the high-speed camera, laser for optical breakdown, and laser for illumination. To achieve high repeatability, the experiments were performed at least 10 times.

\section{Results and Discussion}

The Opto-hydrodynamic phenomena induced during laser pulse-droplet interaction depend on energy absorption characteristics (strength and location of the LIB) inside the droplet. The features of the droplet dynamics during the interaction mainly rely on the energy distribution of the incident laser (transmission, absorption, and reflection) and the droplet properties (optical density and breakdown threshold). Therefore, hydrodynamic phenomena (deformation, fragmentation, and propulsion of fragments) can be regulated by altering either laser energy or droplet properties.

%\subsection*{Global observation}
When the laser pulse is focused inside the droplet, some part of the energy is transmitted, and the droplet absorbs the rest. A part of absorbed energy creates plasma and causes the droplet to expand and fragment concurrently, and the rest of the energy is taken away by the shock wave. We controlled the location and mode of optical breakdown, which dictates the direction of sheet or propulsion of fragments. It is found that the mode of optical breakdown changes with fluid properties.
Irrespective of the mode of breakdown, the breakup of the target droplet is effective when most of the absorbed energy is utilized to break the central droplet. The breakup of the surrounding droplets depends on the combined effect of shock wave and central droplet fragments, and it is efficient when there is a trade-off between these two parameters. Various time scales relevant to the laser-induced breakup of droplet array are 1. Pulse width (6 ns), 2. shock wave timescale ($\sim$ 10 $\mu$s), 3. breakup timescale of center droplet ($\sim$ 100 $\mu$s), and 4. breakup timescale of outer droplets ($\sim$ 1 ms). The morphology of both the inner and outer droplets remains nearly unaffected through the laser pulse with time in the 6 ns scale. 

In the upcoming sections, we discuss how the laser energy, mode of breakdown, and droplet properties influence the breakup of the central and surrounding droplets.

\subsection{Modes of atomization}

\begin{figure*}[tb]
\centering
\includegraphics[width=.85\linewidth]{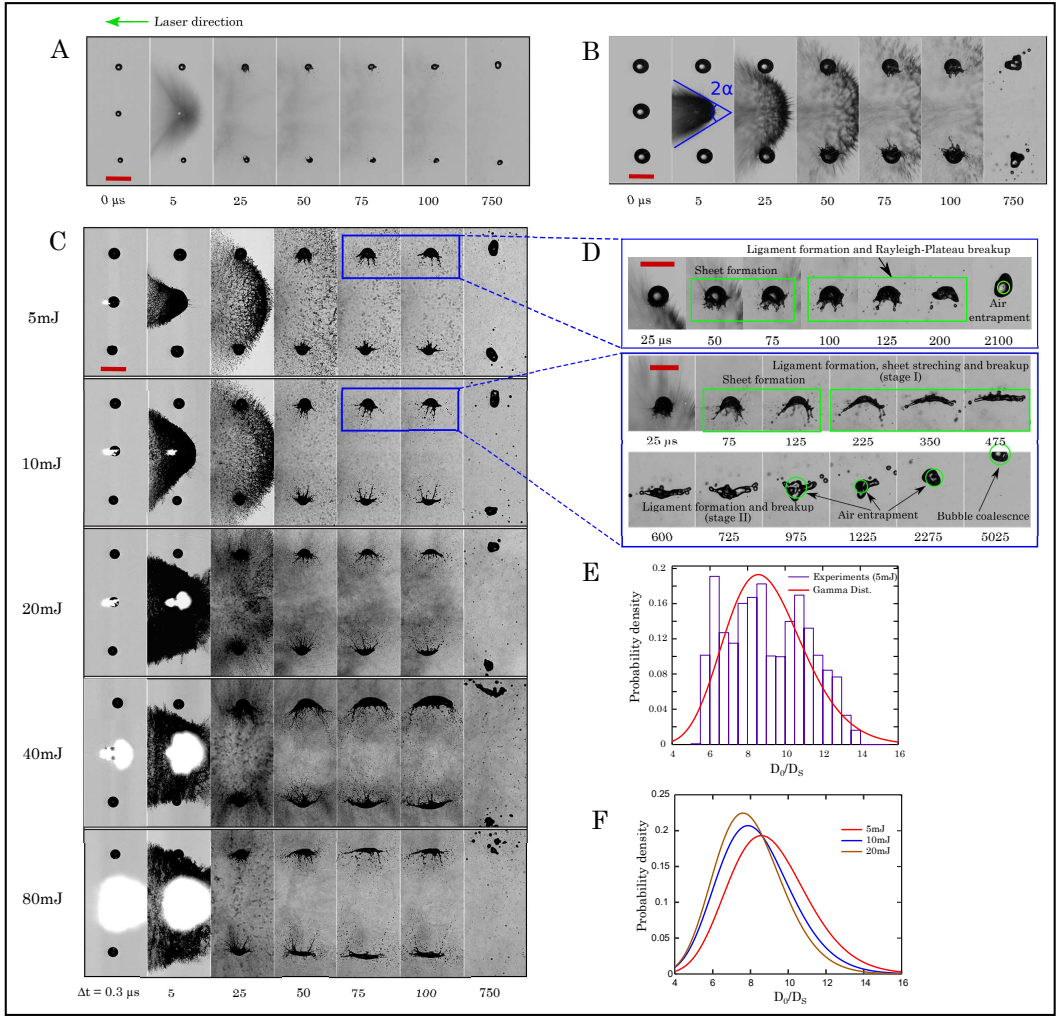}
\caption{Atomization of an array of water droplets corresponding to droplets of  (A) 200 $\mu$m (5 mJ), (B) 600 $\mu$m (5 mJ), and (C) 400 $\mu$m ( 5 mJ to 80 mJ). (D) High magnification images of the breakup of upper droplet corresponding to 5 mJ and 10 mJ pulse energy. (E) Normalized secondary droplet size distribution of water corresponding to 5 mJ pulse energy ( D$_s$ and D$_0$ are the secondary and parent droplet diameters respectively) (F) Comparison of the secondary droplet size distribution for different pulse energies (5 mJ, 10 mJ, and 20 mJ) for the first stage of droplet array breakup. Images shown in (A) and (B) are captured using high-speed imaging while images in (C) correspond to single-shot and high-speed imaging. The scale bar represents 1 mm.}
\label{fig:fig1}
\end{figure*}

Fig. \ref{fig:fig1}A and \ref{fig:fig1}B show the fragmentation of the center droplet for 5 mJ laser energy corresponding to 200 $\mu$m and 600 $\mu$m droplets, respectively. A sequence of breakup events for a 400 $\mu$m droplet for the same incident energy is shown in Fig. \ref{fig:fig1}C. Irrespective of the droplet size, the fragments propel in only one direction (opposite to laser beam direction). A clear distinction in the intensity of breakup is apparent for different-sized droplets (see Fig. \ref{fig:fig1}A-C). The difference in the breakup strength can be attributed to the deposited energy per unit volume available to the droplets for the same incident energy. The influence of acoustic pressure on the trajectories of fragments produced by the atomization process is assumed to be negligible due to the short timescales associated with the dynamics of fragments ($<$ 100 $\mu$s).

With the increment in incident energy (From 5 mJ to 80 mJ), the breakup pattern of the center droplet progressively changes, and the orientation of fragments begins to shift from one direction to all directions (see Fig. \ref{fig:fig1}C at $\Delta$t = 25 $\mu$s). It is witnessed that for 5 mJ and 10 mJ laser energies, a single breakdown occurs towards the shadow side of the droplet (see Fig. \ref{fig:fig4}C). The breakdown prompts the droplet to open from shadow side and results in propulsion of fragments in the opposite direction (illumination side). In contrast, for 40 mJ and 80 mJ energies, multiple breakdowns occur (equally probable in shadow side, droplet center, and illumination side), causing the droplet to open from multiple locations, which in turn disperses the fragments in all directions. In the case of 20 mJ, there is a likelihood of both single and multiple breakdowns, and therefore, it appears to be the transition energy from single-mode to multi-mode breakdown.

For moderately low incident energy (5 mJ), the fragments from the center droplet interact with the surrounding droplets at an angle, $\alpha \approx 35.5^o$. Here, $\alpha$ denotes the divergence angle of the fragments (see Fig. \ref{fig:fig1}B). The uncertainty in the measurement of $\alpha$ is below 2.5\%. The impact of fragments on the surrounding droplets induces minor breakup ensued by shape oscillations. These shape oscillations do not cause a further breakup of the surrounding droplets (upper and lower droplets). It is apparent from Fig. \ref{fig:fig1}C and high magnification images displayed in Fig. \ref{fig:fig1}D that only a small portion of the central droplet fragments interact with surrounding droplets while the larger mass of the fragments propels towards the illumination side. As soon as the fragments strike the surrounding droplets, a thin sheet forms that quickly collapses and, in turn, traps air (Fig. \ref{fig:fig1}D). Meanwhile, the thin sheet develops a rim and consequently leads to the creation and growth of ligaments or threads. These ligaments become unstable and eventually experience breakup due to Rayleigh-Plateau instability \cite{Rao_2020,Chaitanya_Kumar_Rao_2020,Rao_2018_2}. The experimental ligament breakup time ($\sim$ 75 $\mu$s) is consistent with the capillary timescale ($\tau_c = \sqrt{\frac{\rho \times d_l{^3}}{\sigma}}$ = 69 $\mu$s), where $\rho$ and $\sigma$ are the density and surface tension of liquid, and d$_l$ is the ligament diameter. 
When the laser energy is increased to 10 mJ, $\alpha$ is observed to increase ($\sim 41.5^o$). Two stages of breakup are observed; the first stage represents thin sheet formation followed by ligament development and breakup, while the second stage signifies sheet stretching and resulting ligament growth and breakup. Similar to 5 mJ, air entrapment is also witnessed in the case of 10 mJ due to the initial impact of fragments onto the surrounding droplets. However, multiple air bubbles were seen to be trapped inside the surrounding droplets ($\Delta$t = 2250 $\mu$s) that consequently undergo coalescence ($\Delta$t = 5025 $\mu$s). With an additional increase in laser energy (20 mJ), $\alpha$ is observed to decrease ($\sim 23^o$ ) due to breakdown mode transition (Fig. \ref{fig:fig1}C and Movie S1). Therefore, the fragments now collide with the surrounding droplets more evenly than lower laser energies (5 mJ and 10 mJ). The breakup associated with 20 mJ energy is also accompanied by two stages, although no air entrapment is noticed in the residual surrounding droplets. Since the sheet formed due to the impact of fragments keeps expanding, the likelihood of air entrapment inside the surrounding droplet is marginal. For the laser energies greater than 20 mJ, no air entrapment is witnessed.

At higher laser energies (40 mJ and 80 mJ), the fragments collide more symmetrically and evenly with the surrounding droplets since $\alpha \approx 0^o$ due to multi-mode breakdown. However, an additional stage of breakup (third stage) is witnessed, where the expanding sheet becomes substantially thin and subsequently collapses and undergoes further breakup due to significant shape oscillations. This additional breakup stage is associated with the airflow induced by shock wave (will be discussed in the upcoming section). 

\begin{figure*}[tb]
\centering
\includegraphics[width=0.95\linewidth]{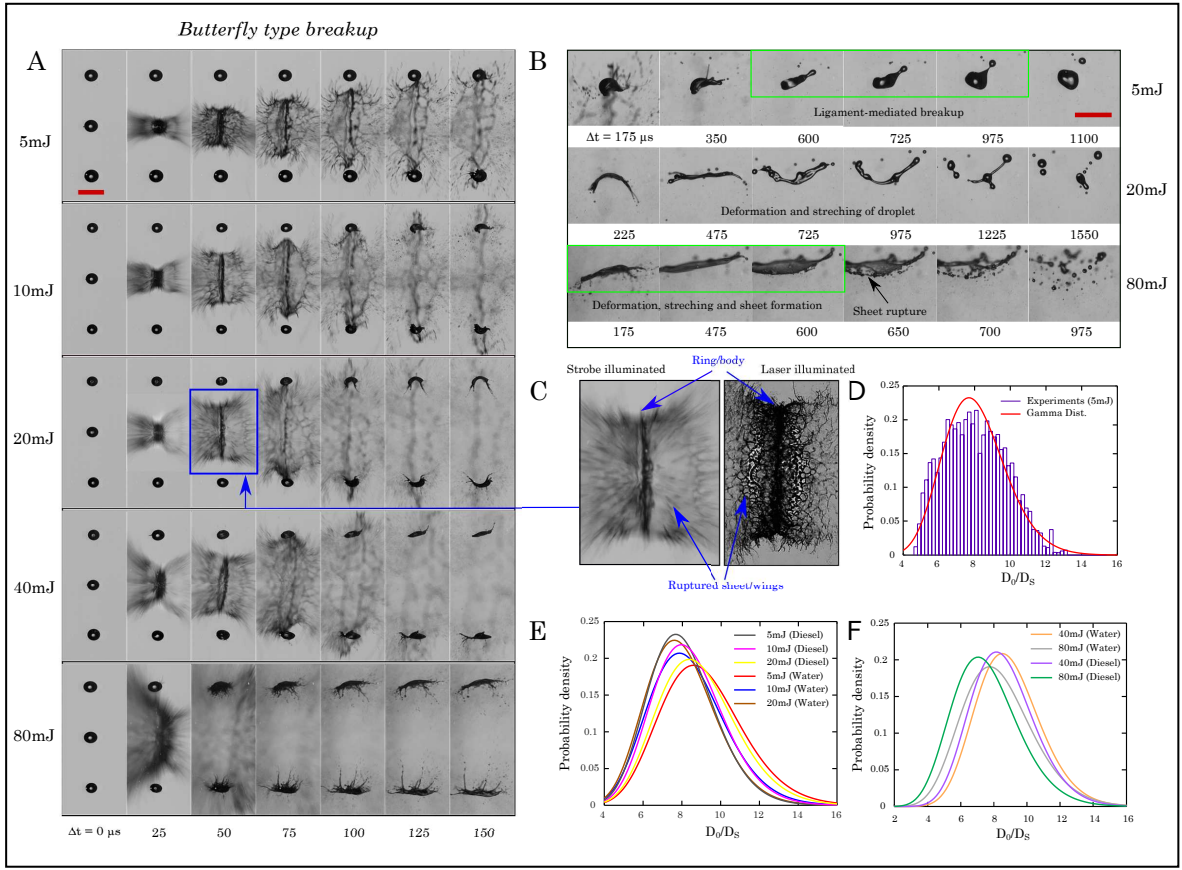}
\caption{(A) High-speed images of breakup in Diesel array droplets corresponding to 400 $\mu$m droplet ( 5 mJ to 80 mJ). (B) High magnification images of breakup of upper droplet corresponding to 5 mJ, 20 mJ, and 80 mJ pulse energy. (C) Butterfly type breakup (strobe and laser-illuminated) in Diesel droplets via laser-induced breakdown of center droplet. (D)  Normalized secondary droplet size distribution of Diesel corresponding to 5 mJ pulse energy (where D$_s$ and D$_0$ are the secondary and parent droplet diameters respectively). (E) Secondary droplet size distribution of Diesel compared to water droplets for different pulse energies (5 mJ, 10 mJ, and 20 mJ) for the first stage of droplet array breakup. (F) Secondary droplet size distribution for multiple stages of droplet breakup for pulse energies of 40 mJ and 80 mJ. The scale bar represents 1 mm.}

\label{fig:fig2}
\end{figure*}

The secondary droplets originating from the surrounding droplets were observed to follow gamma distribution, which is consistent with literature pertinent to secondary droplet formation via ligament-mediated breakup \cite{Kooij_2018,Villermaux_2007}. Moreover, Villermaux \cite{Villermaux_2007} has shown that compared to distributions such as Poisson or log-normal distribution, gamma distributions are more effective for fitting drop-size distribution data. 
The size distribution of secondary droplets for 5 mJ laser energy is shown in Fig. \ref{fig:fig1}E, and in the present study, a similar distribution is observed for all the laser energies. 
For a single (5 mJ) and two-stage breakup events (10 mJ and 20 mJ), it is noted that the size of the secondary droplets is dictated by whether the breakup is dominated by the impact of fragments or ligament-mediated pinch-off. The secondary droplets formed primarily due to the impact of fragments are usually smaller than the droplets generated via ligament-mediated breakup (Fig. \ref{fig:fig1}F). Compared to all other laser energies, it is found that the breakup of the center droplet is more efficient for 20 mJ, and a further increase in energy results in over-killing (i.e. pulse-energy exceeds beyond what is required for efficient droplet fragmentation).

The breakup modes discussed above (for water) are compared with a practical oil such that its breakdown threshold ($\sim 10^{10}$ W/cm$^2$) is sufficiently lower than water ($\sim 10^{11}$ W/cm$^2)$. In this work, we have used Diesel as a representative transportation fuel, which has wide applications in several combustion engines and where drop-drop interactions are dominant. 
%In a combustion engine, it is important to obtain efficient atomization to achieve better combustion efficiency and reduced pollutant emissions such as carbon monoxide, unburnt hydrocarbons, and NOx emissions \cite{Lefebvre_2010}.
Unlike water, due to Diesel's relatively lower breakdown threshold, multiple breakdowns occur even at low laser energy (5 mJ), and interestingly, a new type of breakup is observed, which resembles a butterfly (Fig. \ref{fig:fig2}A, $\Delta$ t=50 $\mu$s). In the butterfly type breakup, the ring represents the body of the butterfly (see Fig. \ref{fig:fig2}A, 40 mJ, 50 $\mu$s) while the wings resemble a thin sheet (Fig. \ref{fig:fig2}C and Movie S2). A laser-illuminated high-speed image of the breakup event indicates that the sheet is fragmented and is composed of a network of holes. For 5 mJ to 20 mJ, the ring radially expands and strikes the surrounding droplets head-on. However, for 40 mJ and 80 mJ, the ring collides with the edge of surrounding droplets due to potential over-killing of breakdown zone. This over-killing causes an imbalance in the energy deposition at multiple locations, which dictates the direction of the ring. Although not as striking as in Diesel, this butterfly-like breakup is also seen for water droplets at the laser energies of 40 mJ and 80 mJ (Fig. \ref{fig:fig1}C).

Fig. \ref{fig:fig2}B shows the high magnification images of upper droplets corresponding to 5 mJ, 20 mJ, and 80 mJ laser energies. In contrast to water, initial thin sheet formation does not occur in Diesel, probably due to its relatively lower surface tension and higher viscosity (Fig \ref{fig:fig2}B, 5mJ). Soon after the impact of fragments, the droplet deforms and stretches, and subsequently undergoes Rayleigh-Plateau breakup. The experimental ligament breakup time ($\sim 25 \mu s$) matches well with the capillary time ($\sim 20 \mu s$). It is evident that, unlike water, the breakup in Diesel is essentially due to ligament-mediated breakup (for 5 mJ to 20 mJ), resulting in a comparatively larger size of secondary droplets. For 20 mJ, the surrounding droplet significantly deforms, flattens, and ultimately fragments into several secondary droplets (via ligament-mediated breakup) following shape oscillations. 

%This ligament-mediated breakup of the deformed droplet leads to a bi-modal droplet size distribution where the small-sized secondary droplets represent satellite droplets. 

Beyond 20 mJ laser energy, the influence of shock wave induced airflow becomes predominant. Accordingly, instead of a ligament-mediated breakup, the airflow ruptures the stretching sheet resulting in the multiple sized secondary droplets (Fig. \ref{fig:fig2}B). Similar to water, the secondary droplets follow gamma size distribution (Fig. \ref{fig:fig2}D) since the breakup is principally ligament-mediated, especially for 5 mJ to 20 mJ (Fig. \ref{fig:fig2}E). A clear distinction in the trend of droplet size distribution is noticed between water and Diesel for laser energies ranging from 5 mJ to 20 mJ (Fig. \ref{fig:fig2}E). In the case of water, as the laser energy is raised, the secondary droplet size also increases due to the dominance of ligament-mediated breakup. In contrast, for Diesel, an increase in laser energy leads to a reduction in the secondary droplet size as the ligament-mediated breakup becomes less imperative. At high laser energy (for 40 mJ and 80 mJ), the trend for secondary droplet sizes becomes identical for both water and Diesel due to the dominance of shock-induced airflow (Fig. \ref{fig:fig2}F).

It is important to note that the breakup dynamics (such as ligament-mediated breakup and sheet breakup) are significantly dominated by the relative influence of energy deposited in the center droplet, the energy for the breakup of center droplet, and shock-induced airflow. In the following section, we will address causes of ligament-mediated breakup, sheet breakup, over-killing of breakdown zone, and the influence of shock wave on the breakup of the surrounding droplets.

\subsection{Dynamics of Shock wave}

\begin{figure*}%[tbhp]
\centering
\includegraphics[width=0.95\linewidth]{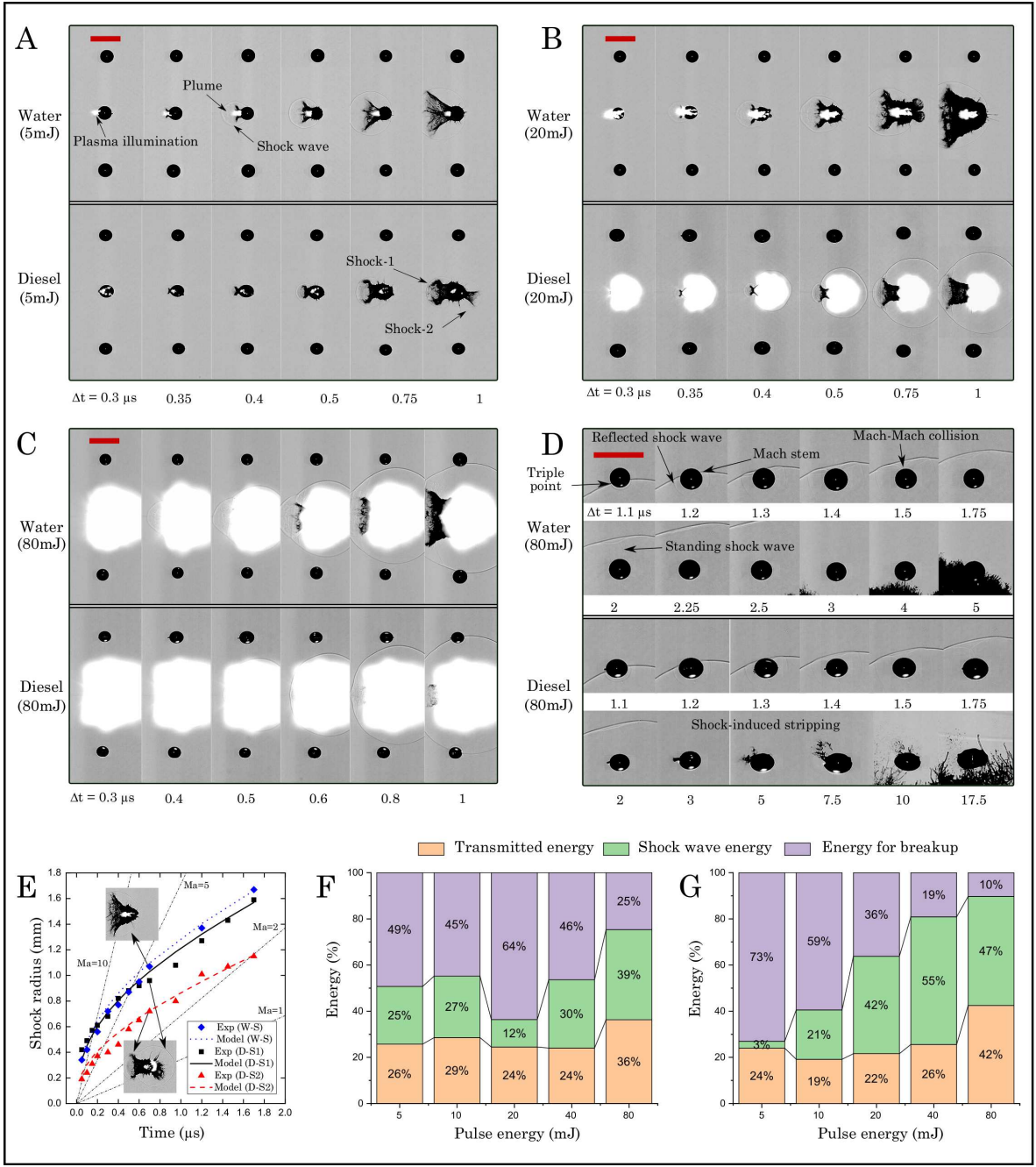}
\caption{Single-shot images corresponding to evolution of early dynamics of the shock wave for water and Diesel center droplet via laser-induced breakdown for (A) 5mJ, (B) 20mJ, and (C) 80 mJ pulse energy. (D) Interaction of shock wave with water and Diesel upper droplet for 80 mJ pulse energy. (E) Comparison of experimental and theoretical evolution of shock wave radius with time for water (single-mode breakdown) and Diesel (multi-mode breakdown) droplets. Distribution of transmitted energy, shock wave energy, and energy for breakup for different laser pulse energies corresponding to (F) Water and (G) Diesel. The scale bar represents 1 mm.}
\label{fig:fig3}
\end{figure*}

%During early nanoseconds, when the laser is focused inside the droplet, the interaction of photons leads to a multi-photon ionization of atoms and molecules in the liquid droplet \cite{Phuoc_2006}. Seed electrons further absorb energy from incoming photons and increase their kinetic energy.  The high-energy electrons then collide with atoms and molecules and ionize them. The cascading of the absorption of photons via electrons and the further ionization of atoms and molecules leads to an optical breakdown. 
The focusing of laser pulse inside the droplet results in optical breakdown. The laser-induced breakdown triggers the radial outward movement of the plasma and shock wave (multiple shock waves in case of multi-mode breakdown). The plasma and shock wave move together up to a few hundred nanoseconds (approximately between 200-400 ns, depending on the pulse energy) (Fig. \ref{fig:fig3}A). Subsequently, the shock wave detaches from the plasma and takes away most of the absorbed energy. Therefore, the energy left with plasma causes the breakup of the central droplet, and the generated fragments move radially outward behind the shock wave. 

The shock wave appears at approximately 350 ns following the laser-droplet interaction (Fig. \ref{fig:fig3}A). The shock wave is visible in the images along with the vapor plume and the mist cloud (two-phase mixture of vapor and fragments of drop, noticeable as a gray to black haze). For a short period (up to 1 $\mu$s), the fragments propel just behind the shock wave. 
The white spot in the center is the high-temperature plasma visible for at least 5 $\mu$s. For a water droplet at 5 mJ energy, a single shock wave evolves with time due to a single-mode breakdown (Fig. \ref{fig:fig3}A). For the same energy, multi-mode breakdown occurs in a Diesel droplet, resulting in multiple shock waves and initiation of breakup from both the shadow and illumination side (1 $\mu$s in Fig. \ref{fig:fig3}A). This multi-mode breakdown results in the earlier mentioned butterfly type breakup. It is also evident that the shock wave generated for water is stronger compared to Diesel (Fig. \ref{fig:fig3}A). This signifies that the energy for breakup for water is significantly lower than that of Diesel for low laser energy (5 mJ and 10 mJ). 
For 20 mJ laser energy, compared to water, LIB in Diesel results in a higher shock strength (and therefore higher shock velocity) and lower energy for breakup (lower fragments velocity) (Figs. \ref{fig:fig3}B and S1, and Movies S3 and S4). A similar trend follows for higher laser energies (40 mJ and 80 mJ), where the shock strength is more significant for Diesel compared to water (Fig. \ref{fig:fig3}C). 
 
As discussed earlier, at high laser energy, the shock wave interaction with surrounding droplets is more dominant. When an incident shock wave interacts with a convex surface (surrounding droplet), Mach reflection occurs when the angle of incidence is below a critical value (known as detachment angle) \cite{Sembian_2016}. A triple point comprising an incident shock wave, reflected shock wave, and a Mach stem can be seen in Fig. \ref{fig:fig3}D ($\Delta$t = 1.1 $\mu$s). The evolution of the reflected shock can be seen between $\Delta$t = 1.1 and 1.5 $\mu$s in Fig. \ref{fig:fig3}D).
 Mach-Mach collision takes place when the incident shock passes the surrounding droplet. Due to the Mach-Mach collision (1.5 $\mu$s), a locally supersonic flow is created, resulting in standing shock waves (2 $\mu$s). The shock-droplet interaction in this study is similar to shock interaction with a solid sphere considering no stripping takes place. Shock-induced stripping is observed for Diesel surrounding droplets, which can be attributed to a high shock strength and lower surface tension (7.5 $\mu$s).

For the breakup of central droplet, only the energy remaining with plasma is valuable. However, for the surrounding droplets, both shock wave and central droplet fragments play a significant role. Therefore, it is essential to determine the shock wave energy for the breakup of both central and surrounding droplets. In this regard, we have quantified the shock wave energy using a model presented by Jones \cite{Jones_1968}. A similar model has been used in various other studies to calculate the shock wave energy after LIB \cite{Gebel_2015,An_2017,Padhi_2020}. Fig. \ref{fig:fig3}E shows the comparison of experimental data (shock radius) with the Jones model. It can be seen from the figure that the Jones model closely matches the experimental data for single (in water) and multi-mode (in Diesel) type breakdowns. The shock wave decelerates very quickly (from Ma$\approx$10 to Ma$\approx$2) within a period of 2 $\mu$s. The experimental uncertainty due to the laser's pulse-to-pulse variation is measured to be less than 3\%. 
Fig. \ref{fig:fig3}F and \ref{fig:fig3}G show the energy balance (transmitted energy, shock wave energy, and energy for breakup) of the optical breakdown in water and Diesel, respectively. In this study, we have neglected the energy losses due to radiation, reflection, and scattering as it has been reported that their combined contribution is less than one percent of the incident energy \cite{Vogel_1999,Joarder_2013}.

In energy balance, the transmitted energy depends on the absorption characteristics of the fluid. It is observed that Diesel absorption is higher compared to water at lower pulse energy due to its relatively lower breakdown threshold. Therefore, the center droplet breakup is more efficient at a lower energy (5 mJ) in Diesel (73 \%) compared to water (49 \%) due to a higher probability of breakdown. For water, central droplet breakup is more efficient (64 \%) at 20mJ laser energy.
As the energy increases (higher than 5 mJ in Diesel and 20 mJ in water), it over-kills the breakdown zone. It is noticed that both absorption characteristics and pulse energy determine the mode of the breakup. If the absorption is higher, like Diesel, the laser pulse-droplet interaction will result in multi-mode breakdown even at lower energies (see Fig. \ref{fig:fig4}G-J), resulting in a butterfly-type breakup. However, if the breakdown threshold is relatively higher (like water), it may result in a single breakdown at three possible locations, i.e., shadow, center, and illumination side of the droplet (see Fig. \ref{fig:fig4}D-F). The location of the LIB determines the direction of movement of the fragments. If the breakdown occurs in the illumination side of the droplet, the fragments will move towards the shadow side (see Fig. \ref{fig:fig4}C). When LIB takes place in the center of the droplet, the fragments move equally in all directions. And when it occurs in the shadow side of droplet, the fragments move towards the illumination side. It is worth noting that we can control and manipulate the breakdown within the droplet, hence the movement of the droplet fragments. However, once the breakdown zone is over-killed, it causes multiple breakdowns (hence multiple shocks) within the droplet and leads to the butterfly type breakup, regardless of the fluid.

Due to this over-kill, the multiple shocks merge within a few hundred nanoseconds and travel as a single intense shock (see Fig. \ref{fig:fig3}B). The potent shock wave takes away most of the absorbed energy, and only limited energy is available to break the center droplet. Since over-killing starts in Diesel from significantly less energy ($>$5 mJ), the shock velocity is faster compared to water; whereas, the fragment velocity is much less than the water (see Figs. \ref{fig:fig3}D and S1). Furthermore, it is observed that irrespective of the droplet properties, after the over-killing of breakdown is reached, shock wave energy increases linearly while energy for breakup decreases linearly (Fig. \ref{fig:fig3}F and \ref{fig:fig3}G). In contrast, the variation in energy for shock wave and breakup is a nonlinear process when there is no over-killing.

The breakup of the surrounding droplets is dependent on the combined effect of shock-induced flow and fragments of the center droplet. It is observed that the breakup is more efficient at higher energies irrespective of the over-killing of the breakdown zone. Therefore, to understand the role of shock-induced flow on the breakup of surrounding droplets, we created a breakdown in significantly small-sized droplets ($\sim$ 20 $\mu$m) such that the influence of fragments on the surrounding droplets is negligible. The creation of breakdown in a small-sized droplet resembles laser-induced breakdown in air \cite{Singh_2019, Bradley_2004,Bak_2015}.

When a significantly smaller central droplet ($\sim$20 $\mu$m) is suspended while keeping the surrounding droplets large enough (600 $\mu$m), the airflow behind the shock wave interacts and deforms the surrounding droplets, indicating that the shock wave has a significant influence on their breakup (Fig. S2). It can be deduced that the shock wave itself can contribute to the substantial deformation and breakup of the surrounding droplets at high laser energy ($>$ 40 mJ) due to significant shape oscillations. This breakup process by shock wave does not expose the droplets to laser irradiation, which has great significance when objects are photosensitive. See supplementary material (Figs. S2 and S3) for further details.

\section{Conclusion}
In summary, by employing a non-intrusive technique, we show that the deformation, breakup, and shock dynamics corresponding to a droplet array can be accurately controlled and manipulated. The change in morphology of droplet array originating from an interplay of incident laser pulse energy (ranging from 5 mJ to 80 mJ) and liquid properties (water and Diesel) is studied in detail. For water, the optimum breakup of the array (both central and surrounding droplets) is achieved at 20 mJ laser pulse energy, beyond which over-killing of the breakdown zone occurs. In the case of Diesel, over-killing begins to take place after 5 mJ energy, and the breakup is effective at different energies for the central (5 mJ) and surrounding droplets (80 mJ). It is observed that irrespective of the droplet properties, after the over-killing of breakdown is reached, shock wave energy increases linearly while energy for breakup decreases linearly with incident laser energy. In contrast, the variation in energy for shock wave and breakup is a nonlinear process when there is no over-killing.
In addition, we report a new butterfly type breakup, which occurs due to multi-mode breakdown. This mode of breakup is found to be an efficient way to deform and fragment the surrounding droplets. Since the present method offers a direct visualization and control of breakup events corresponding to droplet array, it can be further extended to study different photosensitive materials/surfaces (e.g., eye) where a direct interaction of laser with the target is not desirable. The findings of this work on the interaction of laser with the droplet array have strong bearings on various applications such as droplet-droplet interaction in combustion engines, intraocular microsurgery, surface cleaning, to name a few.

%\nocite{*}

\bibliography{apssamp}% Produces the bibliography via BibTeX.

\clearpage 
\appendix
\section{Relative variation of shock and fragments velocity}

Fig. S1 shows the relative dominance of shock velocity over fragments velocity for water and diesel droplets at different laser energies. In case of water, for 5 mJ laser energy, shock velocity is much more dominant compared to the velocity of fragments, which indicates that the shock takes away most of the absorbed energy while significantly low energy is left for the breakup of droplets. A marginal increment in fragment velocity is observed for 10 mJ energy. At 20 mJ energy, the breakup of central droplet is most efficient since larger energy is available for the breakup of the droplet. Similarly, the breakup of the surrounding droplets is also efficient at 20 mJ because the fragments and the shock wave propel at similar velocity. As the laser energy is further increased to 40 mJ and 80 mJ, over-killing of breakdown zone occurs and the shock wave takes away most of the energy.

In case of diesel, since over-killing occurs at energy greater than 5 mJ, a further increase in energy leads to a progressive decrease in fragment velocity because of lower energy left for the breakup of centre droplet (Fig. S1).

\renewcommand{\thefigure}{S1}
\begin{figure}[b]
\centering
\includegraphics[width=0.55\textwidth]{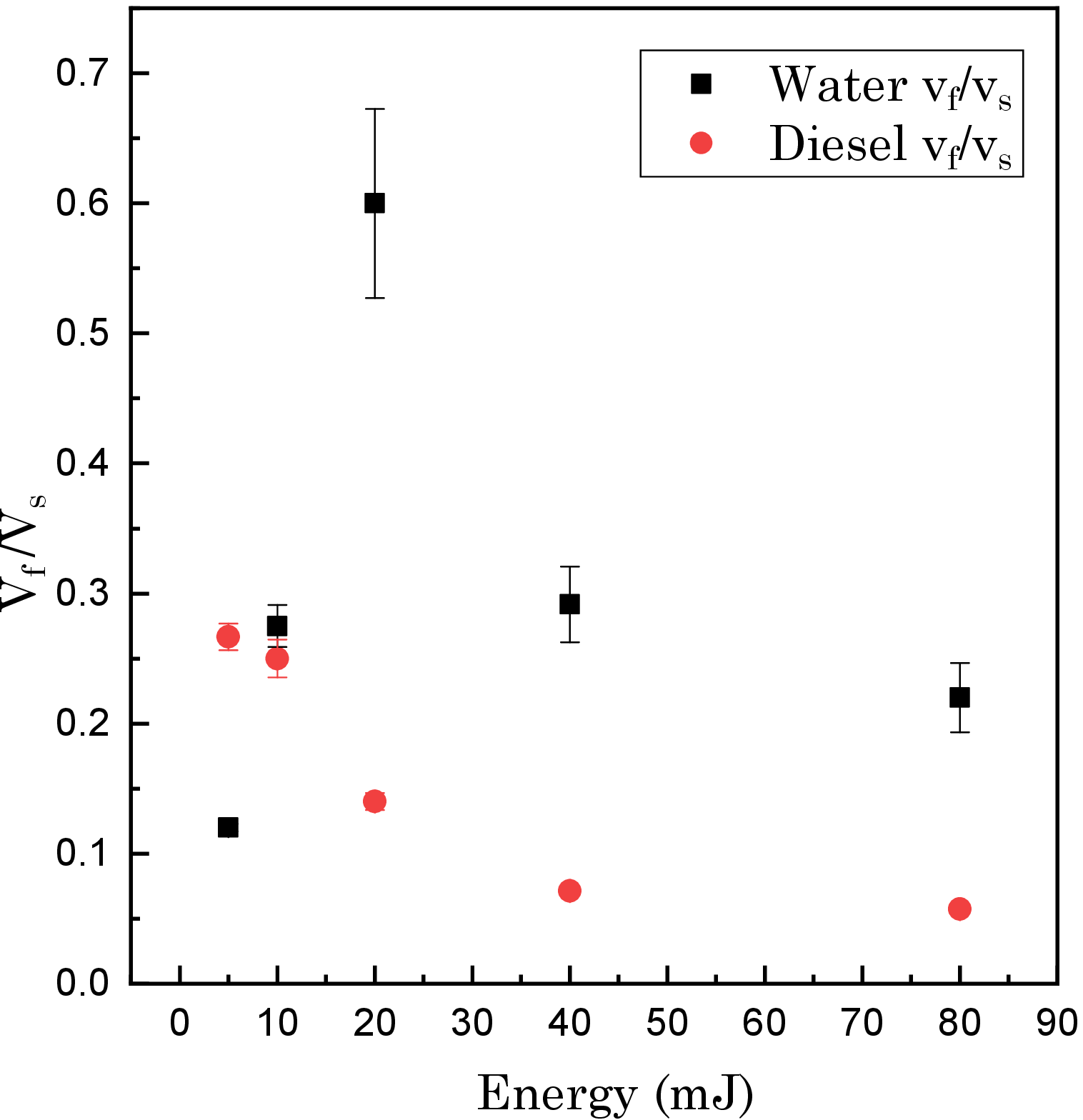}
\caption{The variation of ratio of fragment velocity to shock velocity with the incident laser energy.}

\label{fig:figS1}
\end{figure}

\section{Deformation and breakup of droplet due to shape oscillations}

When an external force is applied to a spherical droplet that is not disturbed by its weight, it oscillates in different resonance modes. The resonance frequency of the corresponding n$^{th}$ resonance mode depends on the physical properties of the liquid and the droplet radius. For a small amplitude oscillation of a spherical droplet immersed in another fluid, a general theoretical expression of the resonance frequency is given by Lamb \cite{lamb1993hydrodynamics},

\begin{equation}
\omega_{n}^{2} = \frac{n(n+1)(n-1)(n+2)}{(n+1)\rho + n \rho^{'}} \frac{\sigma}{R^{3}}
\label{eq:one}
\end{equation}

where $\rho$  and  $\rho^{'}$ are the density of the droplet and the surrounding medium, respectively. $\sigma$ is the surface tension and $R$ is the radius of the droplet. 

The oscillation period obtained from equation B1 for Diesel and water are  968 $\mu$s and 629 $\mu$s, respectively, and it closely matches with experimental results. See supplementary material (Figs. S2 and S3) for further details.

\subsection{Multi-mode resonance}

\renewcommand{\thefigure}{S2}
\begin{figure}[b]
\centering
\includegraphics[width=0.55\textwidth]{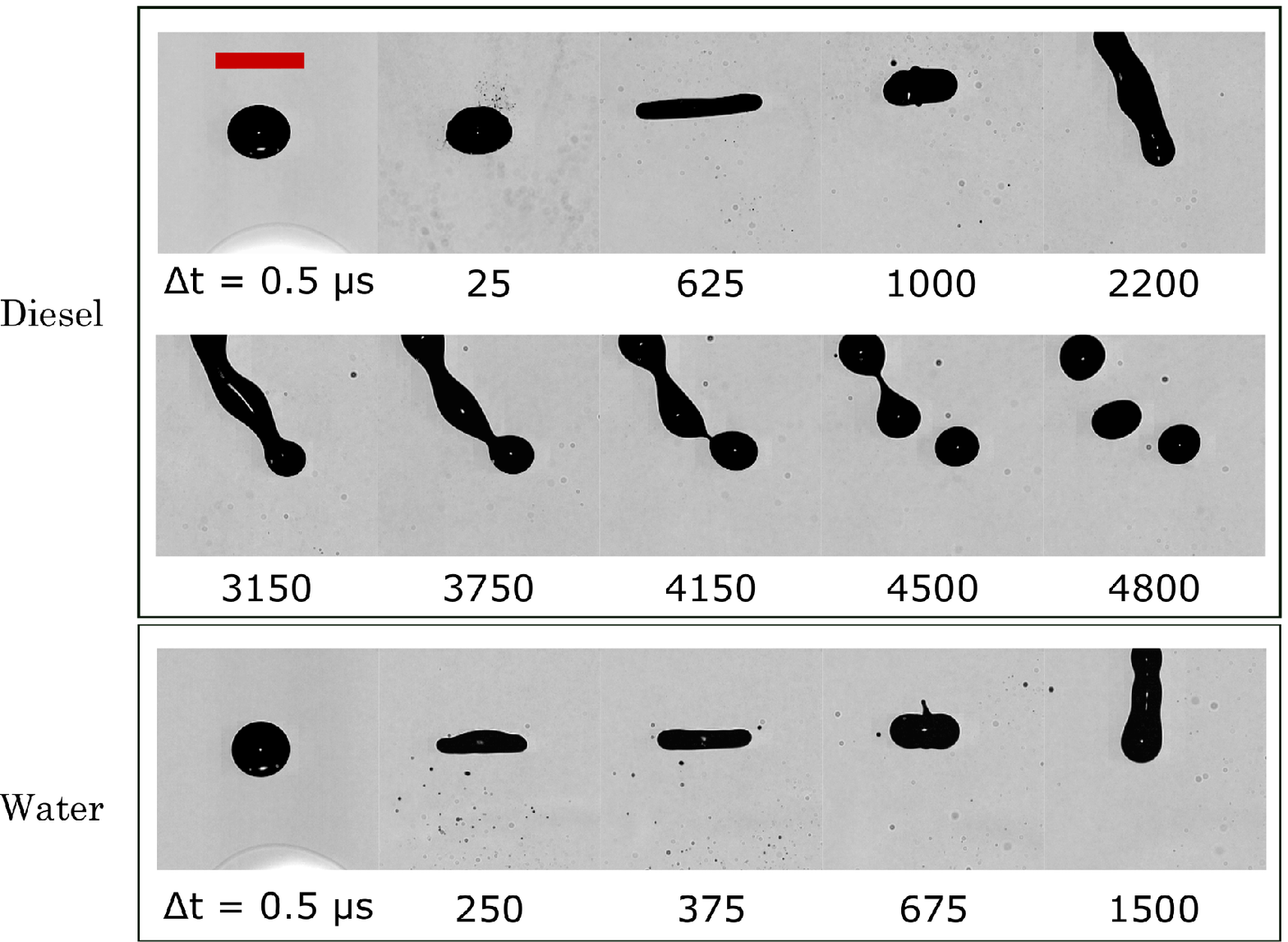}
\caption{Deformation and breakup of surrounding droplets via shape oscillations due to shock-induced airflow for Diesel and water droplets.}
\label{fig:figS2}
\end{figure}

When the shock wave interacts with the surrounding droplet, initially, the Diesel droplet deforms into a disk and begins to resonate ($\Delta$t = 625 $\mu$s in Fig. S2). Subsequently, necking starts to occur on the deformed droplet, resulting in three secondary droplets. Similar deformation and breakup also occurs for water droplets (Fig. S2).

Droplet deformation begins at 10 $\mu$s (in Diesel) and 15 $\mu$s (in water) after the laser pulse is triggered. Fig. S2 (at 1000 $\mu$s in Diesel and 675 $\mu$s in water) shows the point at which the droplet crosses the relaxation point after completing oscillation, indicating an oscillation period of 990 $\mu$s in Diesel and 660 $\mu$s in water. The oscillation period obtained from equation B1 (n=3) for Diesel and water are  968 $\mu$s and 629 $\mu$s, respectively, and it closely matches with experimental results.

\subsection{Single mode resonance}

Single mode breakup is observed in Diesel; however, the probability of its occurrence is considerably low (Fig. S3). Due to the influence of shock induced airflow the droplet first deforms into a disk. However, the deformed droplet does not retract; instead, bag formation occurs in the opposite direction to that of airflow (1350 $\mu$s). If the bag grows and experiences rupture, a ligament is evolved which undergoes Rayleigh-Plateau breakup (3500 $\mu$s).

\renewcommand{\thefigure}{S3}
\begin{figure}[b]
\centering
\includegraphics[width= 0.85\textwidth]{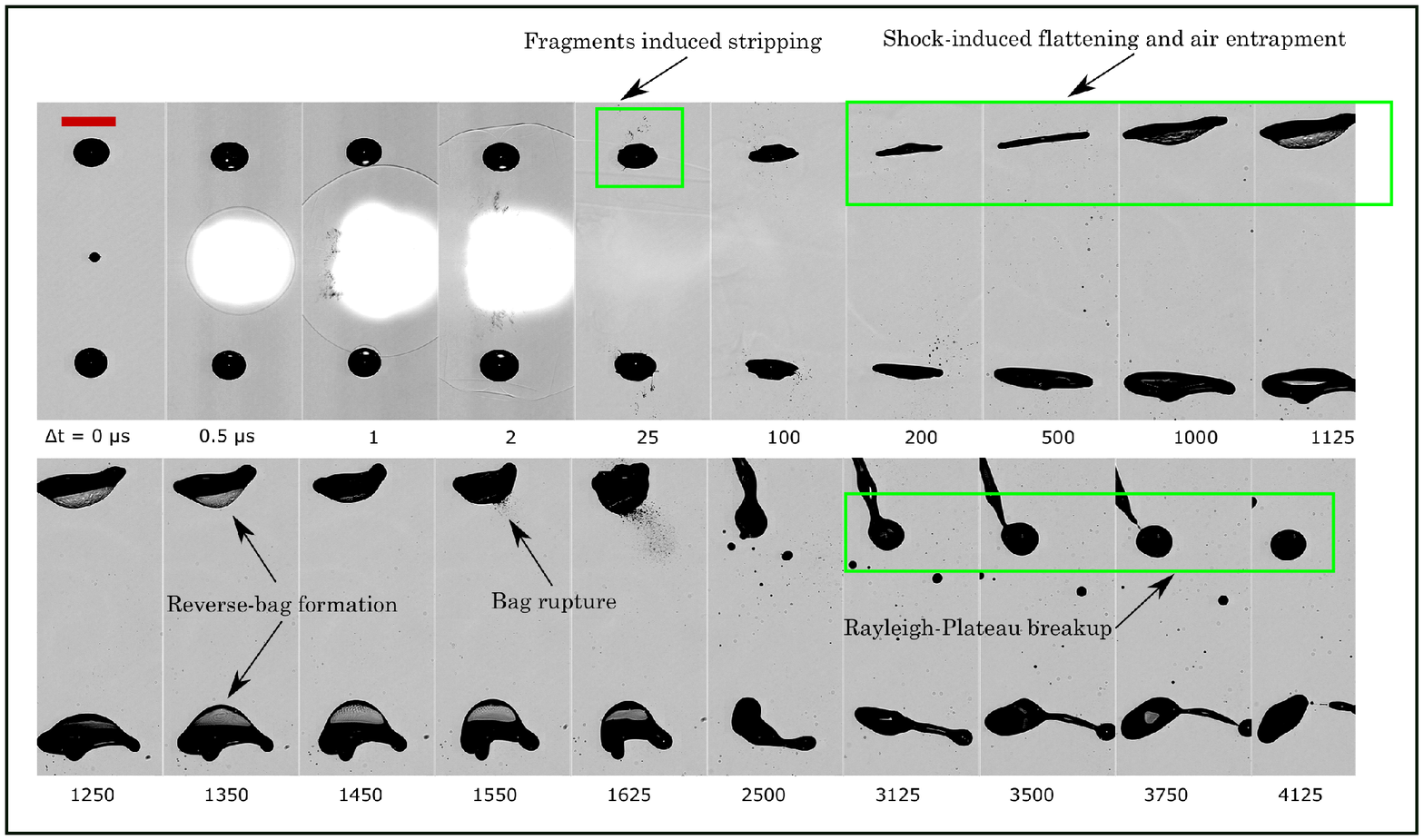}
\caption{Bag formation and rupture in a diesel surrounding droplet due to single-mode shape oscillations}
\label{fig:figS3}
\end{figure}

\section{Supplementary movies}

\setlength{\parindent}{0pt}

\textbf{Movie S1.} {Atomization of an array of water droplets corresponding to 400 $\mu$m droplets for 20 mJ laser energy.}\\
\textbf{Movie S2.} {Atomization of an array of Diesel droplets corresponding to 400 $\mu$m droplets for 20 mJ laser energy.}\\
\textbf{Movie S3.} {Evolution of shock wave and fragments of water droplet via laser-induced breakdown for 20mJ energy.}\\
\textbf{Movie S4.} {Evolution of shock wave and fragments of Diesel droplet via laser-induced breakdown for 20mJ energy.}

\end{document}